\documentclass[aps,showpacs,preprint,amsmath,amssymb]{revtex4}

\usepackage{graphicx} 

\begin{document}
\title{Spontaneous disordering and symmetry breaking in complex plasmas}
\author{S. K. Zhdanov, M. H. Thoma, G. E. Morfill}
\affiliation {Max Planck Institute for Extraterrestrial Physics, 85741 Garching, Germany}
\date{\today}

\begin{abstract}
Spontaneous symmetry breaking is an essential feature of
modern science. We demonstrate that it also plays an important role
in the physics of complex plasmas. Complex plasmas can serve as a
powerful tool for observing and studying discrete types of symmetry
and disordering at the kinetic level that numerous many-body systems
exhibit.
\end{abstract}


\maketitle

\section{Introduction}

In this Letter we address dynamical processes in highly ordered
complex plasmas associated with \emph{spontaneous symmetry
breaking}.

Spontaneous Symmetry Breaking (SSB) plays a crucial role in
elementary particle physics but is also very common in classical
physics \cite{Nambu2009}. It happens whenever the system goes from a
state which has a certain symmetry, e.g. rotational symmetry, into
an ordered state, which does not have this symmetry anymore. In
general, this state not necessarily has to be the ground (vacuum)
state and the transition to the new state may or may not be
associated with a phase transition. For example, in the case of
magnetization the spins point all in one direction (ordered state)
whereas above the Curie temperature there is no preferred direction.
Another example from a mechanical system without phase transition is
a vertical stick which bends under a sufficiently high force from
above to one side breaking the rotational symmetry of the system
without the force.

Different symmetries coexisting in the same phase, and symmetry
transformations escorting phase transitions are widely spread in
nature. For instance, the mechanisms of symmetry breaking are
thought to be inherent in the molecular basis of life
\cite{Goldanskii1989}. SSB is also an important feature of
elementary particle physics \cite{Goldstone1962}. The Universe
itself is believed to have experienced a cascade of
symmetry-breaking phase transitions which  broke the symmetry of the
originally unified interaction giving rise to all known
fundamental forces \cite{Zurek1985,Kibble1985,Bunkov1998}.

Symmetry effects are crucial either in 3D and 2D systems. Chiral
(mirror-isomeric) clusters \cite{Lahav1999}, magic clusters  of a
new symmetry 'frozen-in' by a solid surface \cite{Lai1998}, or
dynamical symmetry breaking by the surface stress anisotropy of a
two-phase monolayer on an elastic substrate \cite{Lu2002} are examples
of the importance of 2D or quasi-2D systems in many
applications.

Low pressure, low temperature plasmas are called \emph{complex
plasmas} if they contain microparticles as an additional
thermodynamically active component. In the size domain of
1-10$~\mu$m (normally used in experiments with complex plasmas)
these particles can be visualized individually, providing hence an
atomistic (kinetic) level of investigations \cite{Thomas1996,Fortov2005}. The
interparticle spacing can be of the order of 0.1-1 mm and
characteristic time-scales are of the order of 0.01-0.1 s. These
unique characteristics allow to investigate the microscopic
mechanism of SSB and phase transitions at the kinetic level.

Common wisdom dictates that symmetry breaking is an inherent
attribute of systems in an active state. Hence these effects are
naturally important in complex plasmas where the \emph{particle
cloud-plasma} feedback mechanisms underlying many dynamical
processes are easy to vitalize. Also in complex plasmas where
different kind of phase transitions exist, e.g. in the
elelectrorheological plasmas \cite{Sutterlin2009}, one can find
examples for classical SSB.  Another option, interesting in many
applications, is the clustering of a new phase which is dissymmetric
with regard to a background symmetry (as an example of fluid phase
separation in binary complex plasmas see \cite{Ivlev2009}).

It is important to mention that the microparticles, collecting
electrons and ions from the plasma background, become charged (most
often negatively \cite{Thomas1996}) and hence should be confined by
external electric fields. The configuration of the confining forces
might deeply affect the geometry and actual structure of the
microparticle cloud. In rf discharge complex plasmas the particles
are self-trapped inside the plasma because of a favorable
configuration of the electric fields \cite{I1994}. One of the
interesting things is the possibility to levitate a monolayer of
particles under gravity conditions. In this case the particle
suspension has a flat practically two dimensional structure. This
is, of course, a very attractive simplification ('from a theoretical
point of view'), significantly lowering the description difficulties.
Below we concentrate mostly on 2D complex plasmas.

Depending on the discharge conditions, the monolayer can have
crystalline or liquid order. 2D configurations of dust particles
either in crystalline  or liquid state were successfully used to
study phase transitions, dynamics of waves and many transport
phenomena in complex plasmas
\cite{Samsonov2004,Knapek2007,Nunomura2005a,Zhdanov2003,Nunomura2006,Nosenko2006,
Nosenko2008}. A symmetry disordering escorting a crystalline-liquid
phase transition has been investigated experimentally in
\cite{Samsonov2004,Knapek2007,Nosenko2008}. Dislocation nucleation
(a 'shear instability') has been reported in
\cite{NosenkoPRL,NosenkoPhylMag}, albeit the importance of SSB
for this phenomenon has not been explained.

The results of these recent experimental observations can not be
properly addressed without a deep understanding of this important
issue. We would like to highlight this in the paper and report on
the physics of spontaneous disordering of a 'cold' plasma crystal,
simulated melting and crystallization process, including associated
defect clusters nucleation, dissociation, and symmetry alternation.
These options are realizable in experimental complex plasmas, and
can be mimicked in simulations, as we demonstrate below.

\section{Spontaneous disordering of a 2D plasma crystal}

It is well known that two broken symmetries distinguish the
crystalline state from the liquid: the broken translational order
and the broken orientational order. In two dimensions for ordinary
crystals it is also well known that even at low temperatures the
translational order is broken by spontaneous disordering mediated by
thermal fluctuations \cite{Berezinskii1971}. As a result, the
fluctuation deflections (disordering) grow with distance and
translational correlations decay (algebraically, see \cite{Landau}).

2D plasma crystals also obey this common rule. The character of
disordering may be deeply affected by the confinement forces,
though. Usually such an 'in-plane' confinement is due to the
bowl-shaped potential well self-maintained inside the discharge
chamber, which to first order is approximately parabolic (see, e.g.
\cite{Konopka,Sheridan2009}), that is $U_{conf}=\frac{1}{2}M\Omega^2
r^2$, where $r$ is the distance, $M$ is the particle mass, and
$\Omega$ is the \emph{confinement parameter} \cite{Zhdanov2003b}.
(The 'out-of-plane' confining forces, controlling the position of
the entire lattice, are normally much stronger; below we consider
the 'pure' 2D-case, assuming, hence, an absolutely stiff
out-of-plane confinement.)

The fluctuation spectra can be calculated in the following manner.
The long-range phonon contribution to the free energy of a 2D system
of particles interacting via the Yukawa potential and confined by a
shallow isotropic parabolic well can be conveniently represented as
\cite{Zhdanov2003}:
\begin{equation}
\label{eq.a} \Delta
U=\frac{M}{2}\sum_\textbf{k}(c_{1}^2|V_\textbf{k}|^2+c_{2}^2|D_\textbf{k}|^2),
\end{equation}
\begin{equation}
\label{eq.2} c_{1,2}^2=c_{tr,l}^2+\delta c^2,~~\delta
c^2=\Omega^2/k^2,
\end{equation}
where $c_{tr,l}$ are the transverse (shear wave) and the
longitudinal (compressional wave) sound speed, and $V_\textbf{k},
D_\textbf{k}$ are the Fourier components of the vorticity $V=curl_z
\textbf{u}$ and the divergency $D=div~\textbf{u}$ of the particle
displacements
$\textbf{u}=\sum_\textbf{k}\textbf{u}_\textbf{k}\exp(i\textbf{k}\textbf{r})$,
and $\textbf{k}$ is the wave vector ($k=|\textbf{k}|$). The
unperturbed crystal is supposed to be hexagonal.

The relationship (\ref{eq.a}) provides  (see, e.g. \cite{Landau})
the probability of the fluctuation $w\sim \exp(-\Delta U/T)$. Next,
using it, we can calculate the averaged fluctuation spectral
intensity per unit mass as
\begin{equation}
\label{eq.3} <|\textbf{u}_{\textbf{k}}|^2>
=\frac{v_T^2}{k^2c_{tr}^2+\Omega^2}+\frac{v_T^2}{k^2c_{l}^2+\Omega^2},~~v_T^2=\frac{T}{M},
\end{equation}
where $v_T$ is the particle thermal velocity.

\begin{figure}[t!]
\includegraphics[width=.35\linewidth,bb = 0 0 185 190]{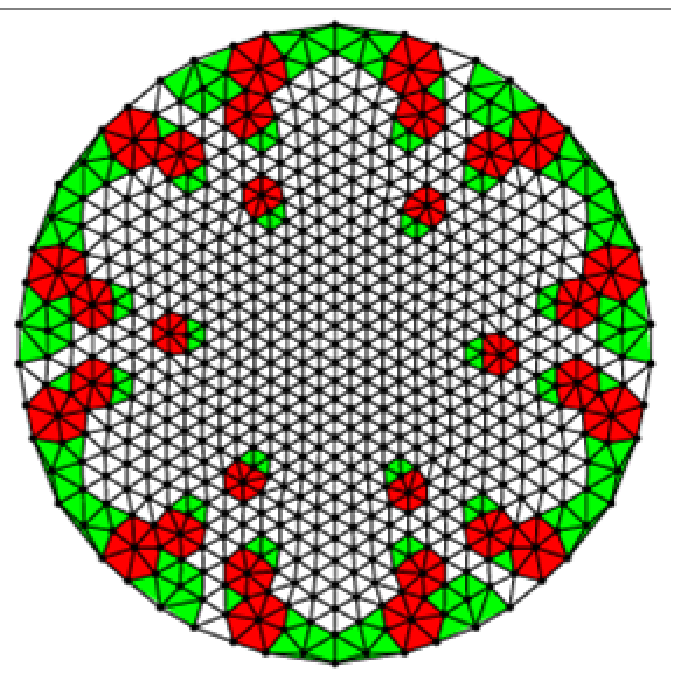}
\includegraphics[width=.55\linewidth,bb = 0 0 249 175]{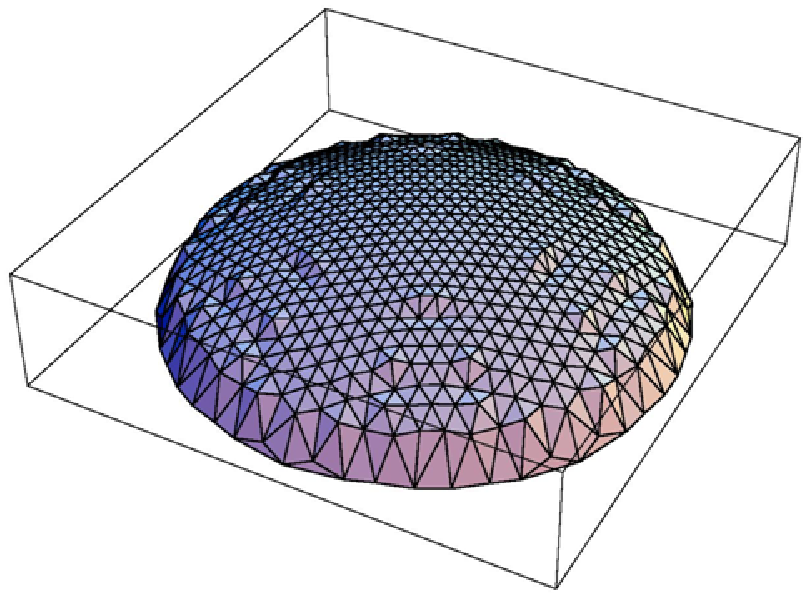}
\caption{\label{fig:Cluster}(color online) Cluster of 721 particles,
interacting via the Yukawa type forces, confined inside the
parabolic well (for simulation details see \cite{Zhdanov2003b}). The
cluster size is $R/a\simeq 40$. Shown are (to the left) the
triangulated particle positions (defect 7- and 5-fold cells are
colored in red and green) and (to the right) the cluster density
profile. On average, the cluster density systematically decreases
toward the edge. The local density maxima (minima) are easily
identified as the positions of 5-fold (7-fold) cells. Note that the
first circular row of the defects happens to appear at $0.46R$ from
the center. This agrees with the theoretical estimate
$\sim(0.44-0.48)R$ obtained by using (\ref{eq.9}). Note also that
the left part and the right part of the simulated cluster image are
perfectly chiral (mirror-symmetric). }
\end{figure}

It is known (see
\cite{Zhdanov2003b,Totsuji2001}) that a lattice layer of a finite size $R$ is stably
confined if roughly:
\begin{equation}
\label{eq.4}\frac{c_{l}^2}{\Omega^2 Ra}\equiv p\simeq const=3-4.
\end{equation}
Here the parameter $p$ stands for an effective number of the nearest
neighbors of any edge particle. Note that according to (\ref{eq.4})
formally $\Omega\rightarrow0$ at $R\rightarrow\infty$.

Without confinement ($\Omega=0$) the fluctuation spectrum
(\ref{eq.3}) apparently diverges $\propto k^{-2}$ at $k\rightarrow
0$, and, as a consequence, in agreement with
\cite{Berezinskii1971,Landau}, the crystal ordering decays
algebraically with the distance $r$, i.e. the density-density
correlation behaves as
\begin{equation}
\label{eq.5}
<\rho(\textbf{r}_1)\rho(\textbf{r}_2)>-\overline{\rho}^2\propto
r^{-n} \cos(\textbf{b}\textbf{r}).
\end{equation}
Here  $\textbf{r}=\textbf{r}_1-\textbf{r}_2$, $\textbf{b}$ is the
vector of the reciprocal lattice, and
$n=\frac{2\pi}{\sqrt{3}}\frac{v_T^2}{c_{tr}^2}$. It is assumed that
$r\gg a$ is large compared to the interparticle separation $a$.

In the experiments  $\Omega\neq 0$ is always finite (though
noticeably small, one or two orders of magnitude less than the
frequency of the local 'caged' oscillations of the individual
particles \cite{Konopka,Sheridan2009}). From (\ref{eq.3})  at
non-vanishing $\Omega$ it immediately follows that the fluctuations
remain finite even at $k\rightarrow 0$.  This absence of a
singularity alters the character of disordering from algebraic
(\ref{eq.5}) to exponential at a scale depending on the confinement
parameter:
\begin{equation}
\label{eq.7} r\sim r_c= c_{l}/\Omega<R.
\end{equation}
It is essential that both asymptotes -- algebraic and exponential --
must be treated as 'near-field' ($r\ll r_c$) and 'far-field' ($r\sim
r_c$) approximations. Hence it would be logical to assume that the
ordering decay alternates with distance from algebraic to
exponential. This is indeed in qualitative agreement with
observations \cite{Quinn1996,Sheridan2008,Nosenko2009}.

Remarkably (\ref{eq.a})-(\ref{eq.3}) are formally similar to the
equations describing director fluctuations in nematic crystals in
the presence of a magnetic field \cite{Gennes1968,Landau}. The
action of the magnetic field is known as suppressing the large-scale
director fluctuations in liquid crystals.

The length scale $\sim r_c$ seems to be of a fundamental importance.
The particles, experiencing a horizontal confinement, are
distributed non-uniformly. The steady-state displacements of the
particles $u$ in the plasma crystal from their ideal locations in a
uniform 2D crystal represent a growing function with distance
$u=r^3/8r_c^2$ \cite{Zhdanov2003b}. The lattice breaks up when
\begin{equation}
\label{eq.7a} \frac{u}{r}=\frac{1}{8}\frac{r^2}{r_c^2}>L,
\end{equation}
where $L$ is the Lindemann parameter. Since $L$=0.16-0.18 (see, e.g.
\cite{Nunomura2006}), it follows that the first row of defects most
probably appears at $r\simeq r_c\sqrt{8L}\simeq (1.1-1.2)r_c$.
Making use of (\ref{eq.4}), (\ref{eq.7}) one can estimate the size
of the domains (or equivalent correlation length) as:
\begin{equation}
\label{eq.9} \frac{r_{cor}}{R}\simeq \sqrt{8pL\frac{a}{R}}.
\end{equation}

The correlation length (\ref{eq.9}) does not depend explicitly on
the temperature. In other words, for purely topological reasons the
big crystal spontaneously splits, assembling an array of
sub-domains, even at zero temperature. The estimated values of $r_{cor}$
agree well with those obtained in the simulation -- see
Fig.~\ref{fig:Cluster}, and in experiments. For instance, it has
been observed in \cite{Sheridan2008,Sheridan2009} that the crystal
orientational order had a power law decay at distances $r/R<1/4$ in
fairly good agreement with $\approx 0.3$ following from
(\ref{eq.9}).

The 'one-plus' correlation length (\ref{eq.9}), unavoidably
introducing a network of sub-domains to a lattice layer, is of
crucial importance, e.g., for observations of the so called
\emph{hexatic state} in the plasma crystals that is still an
outstanding and controversial issue in complex plasma studies
\cite{Quinn1996,Sheridan2008,Sheridan2009}.

\section{Defect clusters in plasma crystals}

\begin{figure}[b!]
\includegraphics[width=.6\linewidth,bb = 14 14 295 228]{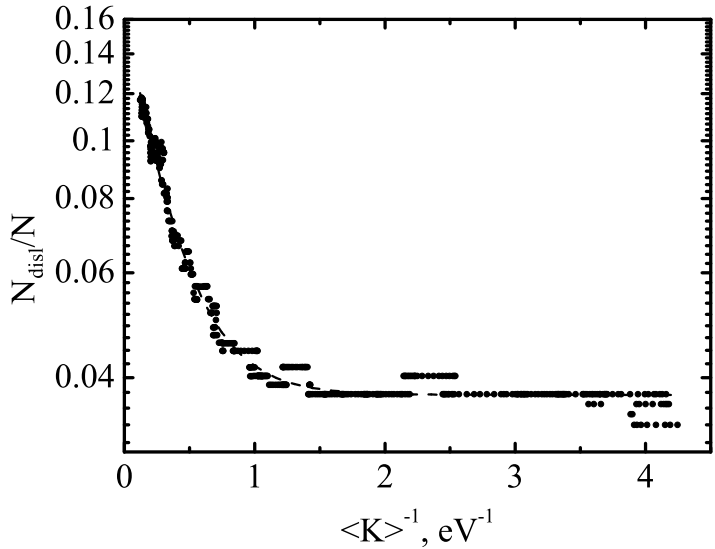}
\caption{\label{fig:Defect}Dislocation density vs. inverse mean
particle kinetic energy. The dashed line is the least squares fit to
the Arrhenius-type dependence $N_{disl}/N=Aexp(-Q/\langle
K\rangle)+B$ plotted with $A=0.12, B=0.037, Q=3.3~\mathrm{eV}$. The
initial exponential drop in the number of dislocations is clearly
seen at $\langle K\rangle^{-1}<2~eV^{-1}$.  Next, at higher $\langle
K\rangle^{-1}$, the  exponential decay is replaced by a power law
decay \cite{Knapek2007,Nosenko2009}: in the figure $N_{disl}/N\approx const$ due to the narrow
range of the inverse mean kinetic energies involved. The simulation
parameters were chosen to match the recrystallization experiment
\cite{Knapek2007}.}
\end{figure}

One of the possible scenarios for melting (recrystallization) in a
2D complex plasma is a precipitous increase (decrease) in the
density of the dislocations and the dislocation aggregates (such as
defect clusters, grain boundaries etc.) \cite{Li1996,Nosenko2009}.
To realize this scenario in simulations, it is desirable to avoid
any aforementioned complications associated with the lattice layer
sectioning 'from the very beginning'. A promising tool in that
sense, allowing to create a defect-free initial lattice layer, is a
hexagonal confinement cell proposed in \cite{Knapek2007}.

We performed a series of simulations that revealed several
peculiarities in symmetry that are worth to mention.

First, the order parameter of the paired defects -- dislocations, --
was systematically lower for 7-fold cells. This is not surprising
actually from a purely geometric point of view because the 5-fold
cell in a pair is more compact.

Second, simulations manifested that not only isolated pairs --
dislocations ($_5$$^7$), but also compact triplets like
($_5$$^7$$_5$), quadruplets ($_5$$^7$$_7$$^5$) etc., or even
elongated defect chains were quite frequent. Actually they
dominantly defined the symmetry of the entire particle suspension.
It would certainly be promising to connect the cluster formation in
ordered complex plasmas \cite{Morfill2009} with the general
percolation process known in many similar applications (see, e.g.,
\cite{Satz1998,Bonn2002}).

Third, in such melted clusters, in agreement with recent
experimental observations \cite{Knapek2007,Nosenko2009}, the defect
density permanently decreased upon cooling. At higher temperatures
in the beginning of the recrystallization process, while the mutual
interparticle collisions were still frequent, the defect density
dropped exponentially. Then, at lower temperatures, the decay rate
significantly slowed down (see Fig.\ref{fig:Defect}).

A sharp drop in the defect numbers followed by a quasi-saturation
resembles the well-known situation \cite{Haenggi1990} in which both
thermal activation and tunneling events occur. Hence, by analogy,
the fact that in our case the system of defects behaves in a similar
way could be naturally explained by an annihilation scenario which
is presumably of the \emph{dissipative tunneling} type
\cite{Haenggi1990,Hofmann1993} at lower mean kinetic energies.

\begin{figure}[b!]
\includegraphics[width=.28\linewidth,bb = 90 11 374 287]{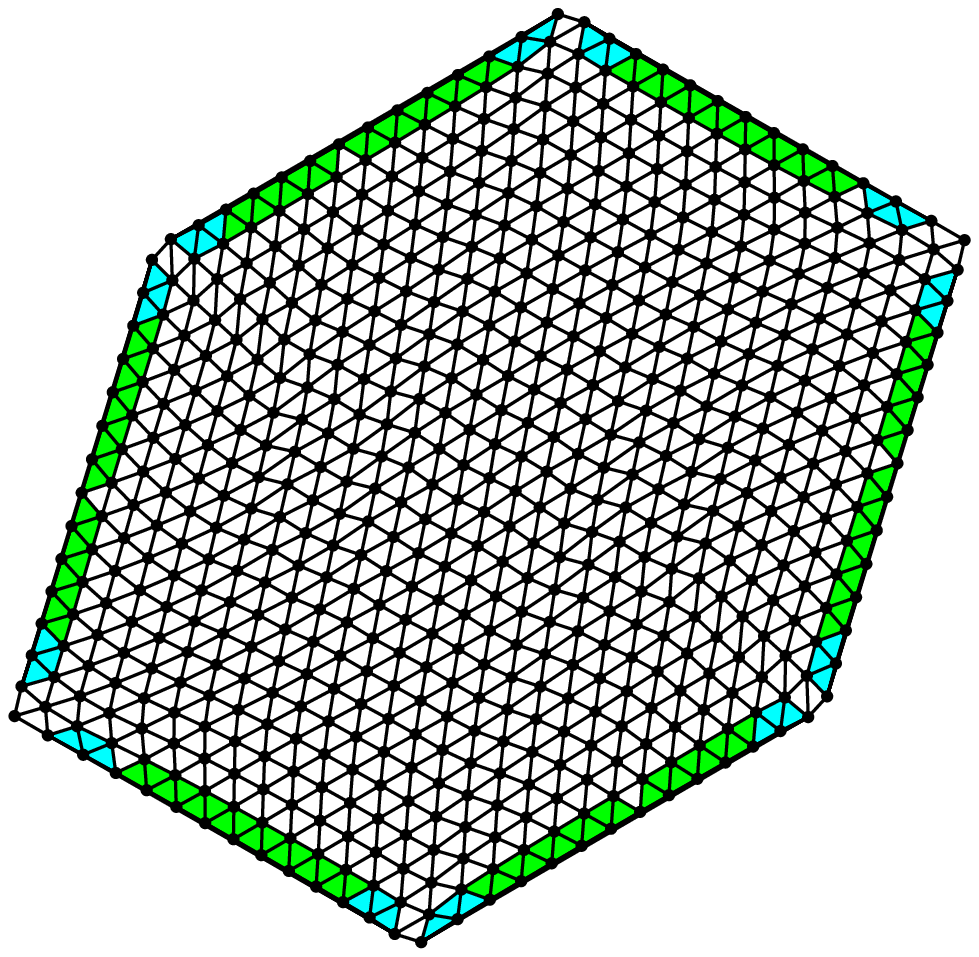}
\includegraphics[width=.28\linewidth,bb = 90 11 374 287]{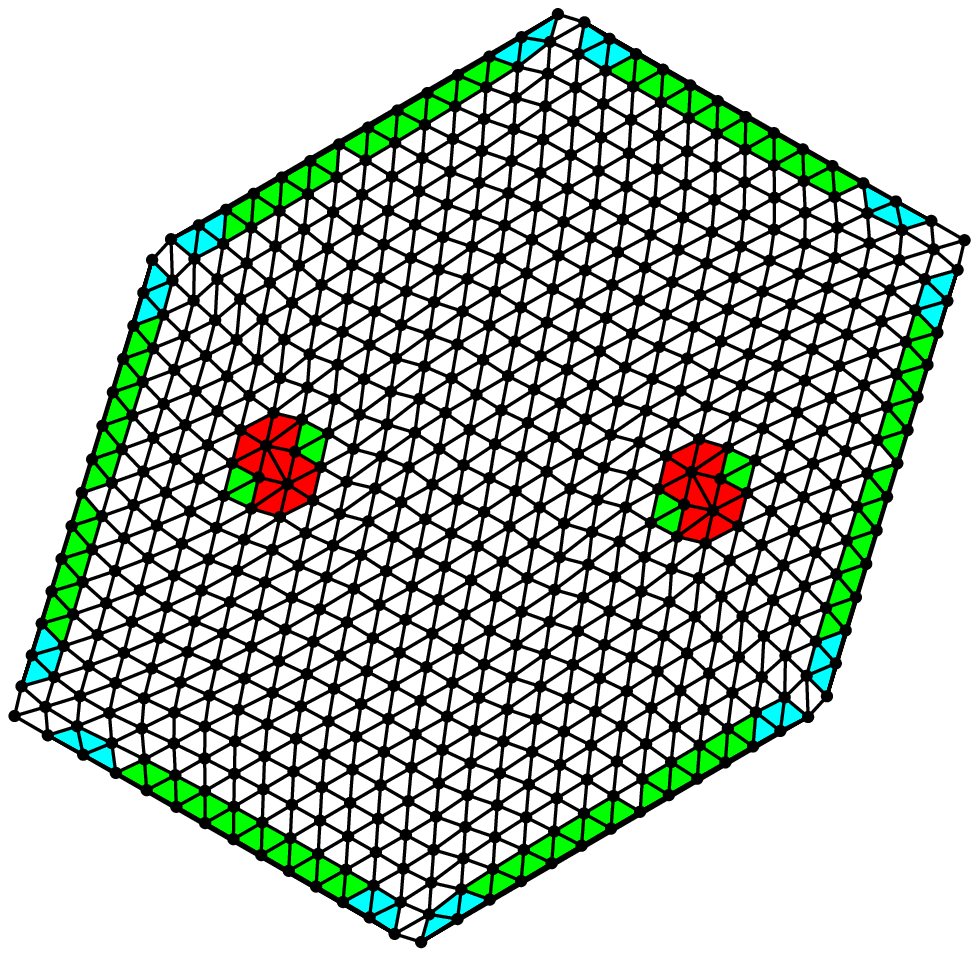}
\includegraphics[width=.28\linewidth,bb = 90 11 374 287]{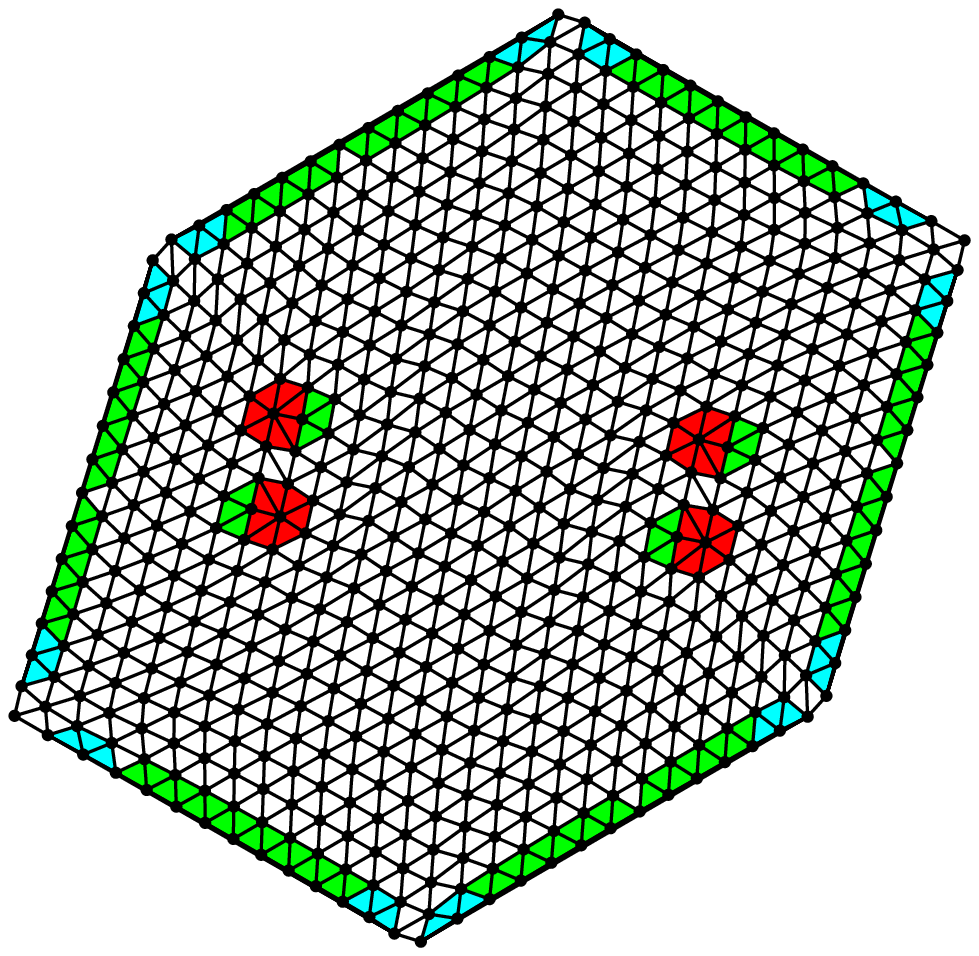}
\caption{\label{fig:Nucleation}Dislocation nucleation affected by a
simple shear strain induced externally. The hexagonal simulation
cell contains 631 interacting particles. The forces, deforming the
cluster, are applied from left to right on top of the cell, and from
right to left on bottom. The central horizontal part of the cell is
kept non-deformed. The deformation rate is
$\tau\dot{\varepsilon}=0.05$, where $\tau$ is the Coulomb time-scale
\cite{Zhdanov2003}, $\varepsilon$ is the shear strain
\cite{Kittel1976}. Three panels show the triangulated particle
positions for the following time instants (from left to right): shortly before
nucleation starts, immediately after nucleation, shortly after
dissociation of the dislocation pair. The newly nucleated
dislocations glide along the lines with the maximal stress. }
\end{figure}

\section{Complexity of nucleation kinetics}

Nucleation of dislocations is another important example of
spontaneous symmetry breaking on a scale of elementary cells.

Whatever the melting scenario would be true, still there would
remain a question what mechanism explains nucleation of the primary
dislocation clusters. Recently this issue has been studied
experimentally: Spontaneous nucleation of the edge-dislocation pairs
(followed by their dissociation) has been successfully observed at
the kinetic level in the experiments with plasma crystals
\cite{NosenkoPRL}. Since the Burgers vector of the entire lattice is
kept constant (e.g. zero) spontaneously created dislocations must be
paired forming defect quadruplets of the type ($_5$$^7$$_7$$^5$).
(The Burgers vector characterizes the magnitude and direction of the
crystalline lattice distortion by a dislocation \cite{Kittel1976}.)
These dislocation clusters were created in the lattice locations
where the internal shear stress exceeded a threshold. It has also
been shown that even an elementary act of nucleation is in fact a
multi-scale process consisting of the latent 'pre-phase', prompt
nucleation of a defect cluster, and dissociation of the cluster
followed by the escape of free dislocations \cite{NosenkoPhylMag}.

In the experiments \cite{NosenkoPRL,NosenkoPhylMag} it was suggested
that the stress that finally caused nucleation was affected by the
differential crystal rotation. The exact reason of nucleation,
however, was difficult to determine consistently. In simulations the
nucleation conditions are certainly easier to identify.

To demonstrate nucleation in simulations a 'deformable' hexagonal
cell is used (Fig.\ref{fig:Nucleation}). It confines a 2D cloud of
equally charged particles interacting pairwise via the Yukawa (the
screened Coulomb) force:
\begin{equation}
\label{eq.10} \textbf{f}_{i,j}=\frac{q^2}{R^2}
\frac{\textbf{R}_{i,j}}{R}
(1+\frac{R}{\lambda})\exp(-\frac{R}{\lambda}).
\end{equation}
where $\textbf{R}_{i,j}=\textbf{r}_i-\textbf{r}_j$ is the relative
coordinate and $R=|\textbf{R}_{i,j}|$ is the distance between the
particles $i, j$ with the coordinates $\textbf{r}_i, \textbf{r}_j$;
$q$ is the particle charge and $\lambda$ is the screening length.
The cell design is similar to that applied in \cite{Knapek2007} to
simulate melting and recrystallization process of the plasma
crystal.

The hexagonal simulation cell has the evident advantage of flexible
shape, compared to, e.g., a parabolic cell confinement. Deforming
the boundary of the cell, it is simple to manipulate the particles
in a tractable way. An additional option of variable geometry
enables an opportunity to separate or consolidate pure shear and
simple shear deformation \cite{Kittel1976} if desirable. The strain
rate is controllable during deformation as well.

Fig.~\ref{fig:Nucleation} shows  a simple-sheared particle lattice
layer. At a properly chosen loading rate deformation affects the
\emph{shear instability} that ends up with nucleation of defect
clusters in the bulk of the lattice layer. After a while, when
deformation becomes stronger, the components of the clusters
decoupled and the newly born free dislocations glided away in a
similar manner as the dislocations observed in experiments.

\begin{figure*}
\includegraphics[width=.5\linewidth,bb = 42 41 368 371]{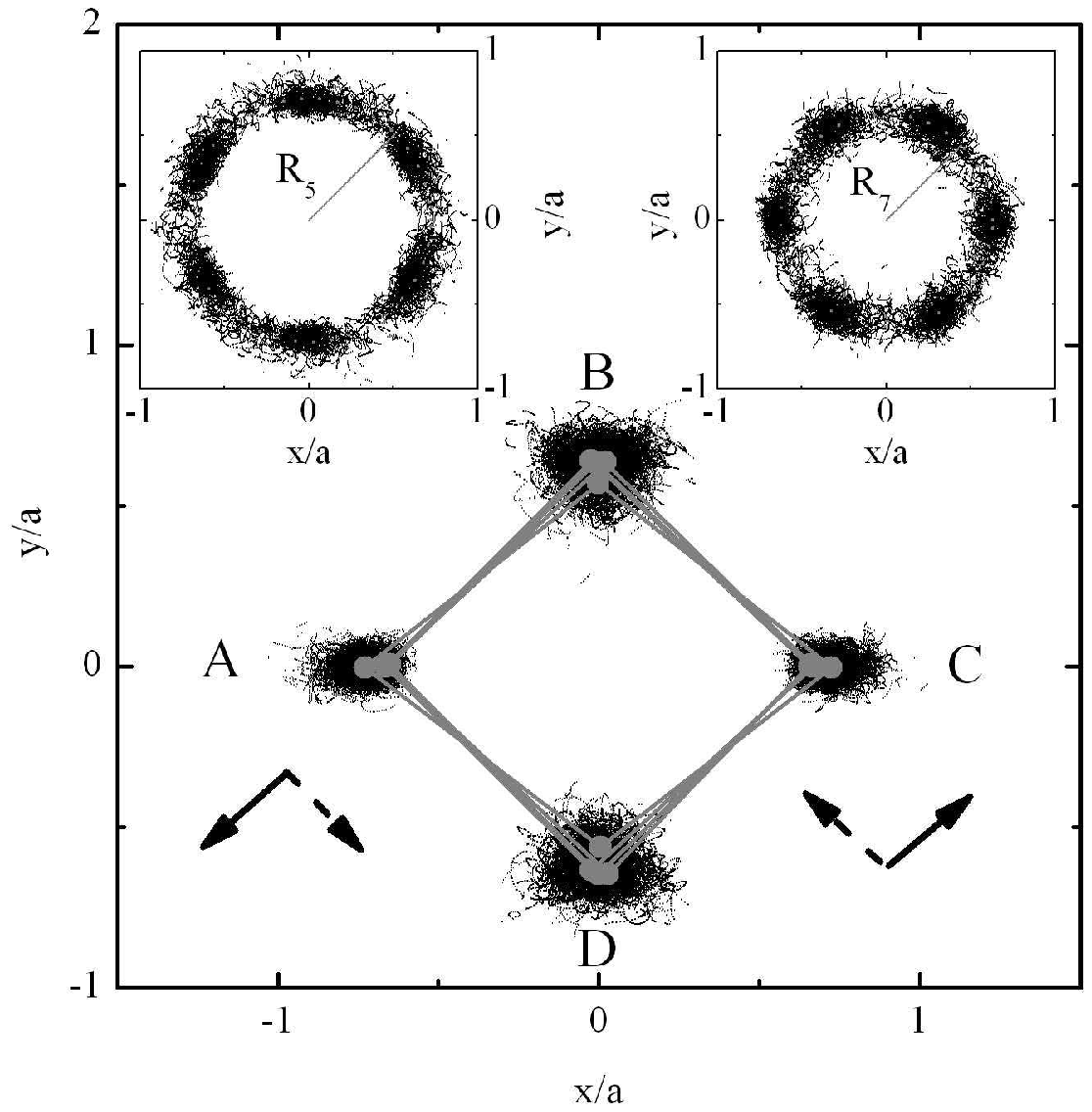}
\includegraphics[width=.5\linewidth,bb = 0 0 607 246]{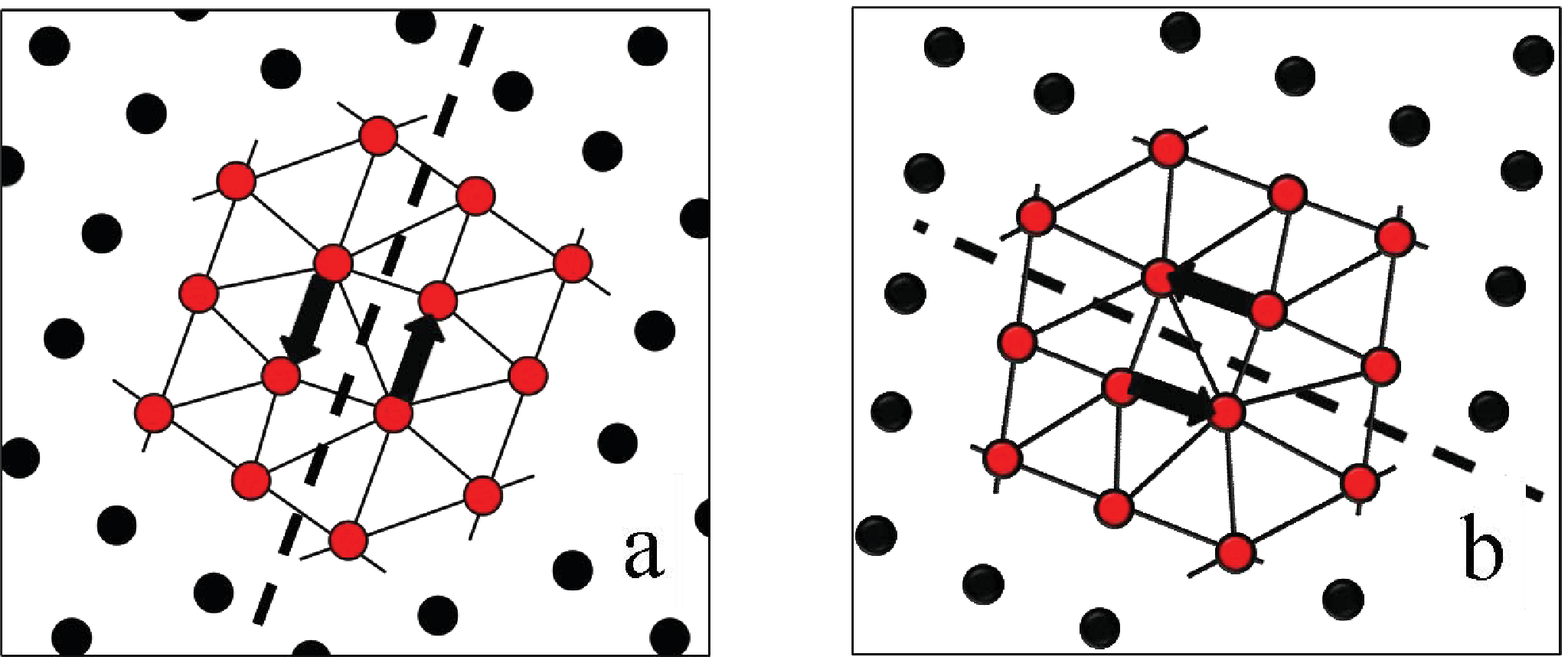}
\caption{\label{fig:topology} Topology of a four-component defect
cluster. (black dots) Position of the centers of the 5- and the
7-fold cells obtained in simulations. The simulation time is $\Delta
t/\tau=120$. The mean kinetic energy of the particles is kept constant
at the level $<K>=7.3\pm 0.2~eV$. The mean size of the cell edges is
$<s>\equiv 1/4<p>=0.96\pm 0.04$, where $<p>$ is the mean cell
perimeter. (As a scaling parameter the lattice constant is used.)
Centers of the 5-fold (A, C) and the 7-fold (B, D) cells compose a
nearly regular compact tetragonal structure. The gray quadrilaterals
represent a few superimposed experimental clusters
\cite{NosenkoPRL,NosenkoPhylMag} shown for comparison. (inserts)
Patterns of orientational preference revealed a visible
'quasi-hexagonal' trend in orientations for (left insert) the 5-fold
and (right insert) the 7-fold components of the clusters. The mean
radii are $R_5=0.721>R_7=0.629$. The bold and the dashed arrows
indicate schematically the Burgers vectors of the individual
dislocations. The components of the clusters (a) and (b) are shown
in the bottom panels. The individual dislocations should have
anti-parallel Burgers vectors since the Burgers vector of the entire
defect group must be zero. This evidently results in the
mirror-isomeric configurations of the escaping dislocations (the
gliding planes are indicated by the dashed lines). }
\end{figure*}

\section{Topology of the dislocation cluster}

Symmetry alternation is of primary importance for understanding
nucleation of dislocation clusters. The compact cluster design is
\emph{magic} in the sense that the \emph{hexagonal} symmetry of the
particle system neatly turns into a nearly \emph{tetratic} symmetry
of the cluster core (like lead turns into gold when touched by the
Philosopher�s Stone), see Fig. \ref{fig:topology}.

Despite an apparent simplicity of the cluster interior -- only four
nearest neighbor particles (marked ABCD in Fig.~\ref{fig:topology}),
the centers of the 5- and 7-fold cells, are in the core, -- to
discover the cluster topology was certainly a challenge
\cite{Maret2005}. In our case the interparticle interaction
potential is of the screened Yukawa type, hence more compact in
contrast to the $\propto r^{-3}$ interaction potential in case of
magnetically interacting super-paramagnetic colloid particles
considered in \cite{Maret2005}. Thus there is a unique opportunity
to verify whether the core topology is universal.

Let us start with a simple model treating the cluster as constituted
of two point-like dislocations, which are set apart at a distance
$r$ and allowed to glide only along two fixed crystallographic
planes separated by one lattice period $a$, so that $r\equiv a/\sin
\varphi$, where $\varphi$ is the angle of mutual orientation of the
cluster components with respect to the gliding plane. The
interaction energy of the point dislocations having the
counter-directed Burgers vectors is
\cite{Peach1950,Peeters1987,LandauLifshitz,Maret2005}:
\begin{equation}
\label{eq.11}
u_{dd}(r)=\frac{2}{\pi\sqrt{3}}Mc_{tr}^2[\ln(\frac{r}{a})+\frac{a^2}{r^2}]+const.
\end{equation}
It has a minimum, a \emph{stable ground state}, at
$r/a=\sqrt{2}=1/\sin \varphi$. It corresponds to $\varphi=45^\circ$,
hence  tetragonal symmetry of the cluster core might be considered
as preferred.

This prediction agrees noticeably well with the results of
simulations of finite clusters: On average in
Fig.~\ref{fig:topology} the edge-to-diagonal angle in the cluster
core is $<\varphi>=42^\circ\pm2^\circ$.

It is worth noting that the cluster core is nearly cyclic. A measure
of it immediately follows from the famous Ptolemy's inequality valid
for any quadrilateral:
\begin{equation}
\label{eq.12} Pt\equiv \frac{s_1s_3+s_2s_4}{d_1d_2}\geq 1,
\end{equation}
where $s_{1,...,4}$ denote the (ordered) sides, and $d_{1,2}$ are
the diagonals of the quadrilateral. Over 80~\% of the recognized
clusters have $1\leq Pt\leq 1.03$ for the simulation results shown in
Fig.~\ref{fig:topology}. For comparison a hexagonal four-side cell
corresponds to $Pt^{hex}=2/\sqrt{3}=1.1547$.

Note also that a stable defect cluster could not be obtained only by
shifting positions of four central particles from a hexagonal
configuration to tetragonal one. Such deformation would be
reversible, hence unstable. A weakly deformed environment, impeding
relaxation of the core particles back to the stable hexagonal
configuration, is indeed a necessary 'lock' making the deformation
plastic, i.e. irreversible.

At sufficiently strong external stress even a stable cluster
dissociates. Whatever is the orientation of the cluster as a whole,
escaping dislocations can glide only along two crystallographic
directions (along Burgers vectors, see Fig.~\ref{fig:topology} (a,
b)). This naturally explains the asymmetry of the escape directions
and chirality of the defect configurations revealed by the newly
nucleated dislocations in experiments \cite{NosenkoPhylMag}.

\section{Summary}

Spontaneous symmetry breaking is a common and inherent feature of
many systems in physics as well as other fields, it plays an
important role, for example, from classical one-component plasmas to
modern string representations \cite{Thoma2008}, from the evolution
of the early universe \cite{Kolb1990} to the dynamics of a wide
variety of small-scale systems \cite{Kibble2007}. Therefore it is
not surprising that SSB is present also in the physics of plasma
crystals. As example we have considered dislocations in plasma
crystals which exhibit a spontaneous disordering, involved in the
process of melting, and form clusters that are caused by a shear
instability and show an interesting topological symmetry.

\section{Acknowledgement}
The authors appreciate valuable discussions with Dr. Ivlev and Dr.
Nosenko.

\end{document}